\documentstyle[12pt]{article}
\input{psfig}
\begin{document} 
\begin{titlepage}	
\begin{center}
{\Large {\bf Scaling of cluster fluctuations in  
two-dimensional $q=5$ and $7$ state Potts models$^\dagger$}} 
\end{center}
\vskip 1.5cm
\centerline{\bf Burcu Ortakaya, Yi\u{g}it G\"{u}nd\"{u}\c{c}, Meral Ayd{\i}n 
and Tar{\i}k \c{C}elik }
\centerline{\bf Hacettepe University, Physics Department,}
\centerline{\bf 06532  Beytepe, Ankara, Turkey }
\vskip 0.5cm

\centerline{\normalsize {\bf Abstract} }

{\small The scaling behavior of fluctuations in cluster size is
studied in $q=5$ and $7$ state Potts models. This quantity exhibits
scaling behavior on small lattices where the scaling of local
operators like energy fluctuations and Binder cumulant can not be
expected.}

\vskip 0.4cm


\vskip 2.0cm

$\dagger${This project is partially supported by Turkish Scientific
and Technical Research Council (T\"{U}B\.{I}TAK) under the project
TBAG-1299.}
\end{titlepage}	


Rigorous theory of finite size scaling, for first-order phase
transitions, put forward by Borg et. al. \cite{Borgs:1990,Borgs:1991}
gave explicit form for the size dependence of various
operators. According to this theory, for periodic boundary conditions,
the operators such as specific heat and Binder cumulant at the
transition temperature can be represented by a polynomial in $ 1/L^d$
where $L$ is the linear lattice size and $d$ is the dimension.  If the
lattice sizes are much larger than the correlation length, the
contribution of the higher order terms are negligible
\cite{Billoire:1995}.  The difficulty arises when the correlation
length is of the order of the lattice size or larger. In this case,
higher order corrections are necessary and deciding the order of the
transition becomes difficult. Even when the large lattices are used,
higher order terms may create difficulties during the fitting process
to the simulation data.  Such difficulties may be reduced by choosing
the operators for which the correction terms play less important
role. A good example for such an operator is the average energy
measured at the infinite lattice transition point. This quantity has
exponentially small correction term which enables one to determine the
infinite lattice critical point with great accuracy
\cite{Borgs:1991,Billoire:1995,Billoire:1992a}.

$\;$ 

In an earlier work \cite{Aydin:1996,Gunduc:1996} on the
two-dimensional Potts model, it is observed that compared to the local
operators such as energy and order parameter, cluster related (global)
operators are more sensitive to the changes in the system going
through a phase transition. Particularly, the average cluster size
distribution may give better indication on the order of the phase
transition at smaller sizes than that for the energy distribution.
The most trustworthy method to test this observation is studying the
finite size scaling behavior of the cluster related operators. In this
work, the finite size behavior of the cluster size fluctuations in
$q$-state Potts model under the light of the finite size scaling
theory, developed for first-order phase transitions in the past decade
[1,2,7,8] will be examined.

$\;$ 

The two-dimensional $q$-state Potts model is known to undergo a
first-order phase transition for $q > 4 $ ~\cite{Baxter:1973}.  While
the correlation length at this transition, $\xi$, is about 10 lattice
sites for $q=10$, ~$\xi$ approaches infinity in the limit as $q
\rightarrow 4$, ~\cite{Buffenoir:1993}.  Even though the $q=7$ Potts
model exhibits strong first-order behavior, correlation length is of
the order of 50 lattice sites. $q=5$ state model has a weak
first-order transition with a correlation length about few thousand in
lattice units.  In various studies on finite size scaling behavior of
the $q$-state Potts model, it has been shown that $q \ge 10$ state
Potts model enters into asymptotic regime for the lattice sizes
$L>40$~~\cite{Billoire:1992a,Billoire:1992b,Lee:1991} (for a review
see \cite{Billoire:1995}).  This makes the observation of scaling
difficult for $q < 10$ in simulation studies by using computationally
feasible lattice sizes
~\cite{Billoire:1992a,Janke:1992,Rummukainen:1993} where higher order
corrections become important. However these studies are based on the
finite size scaling behavior of the local operators such as the energy
cumulants. The aim in this work is to observe the characteristic
behaviours in the system by studying global operators such as
fluctuations in cluster size which may yield information on the
scaling form of the operators and hence on the order of the
transition.

$\;$

The Hamiltonian of the Potts model~\cite{Potts:1952,Wu:1982} is given
by
\begin{equation}
     {-\beta \cal H} = K \sum_{<i,j>} \delta_{\sigma_{i},\sigma_{j}}.
\end{equation}
Here $K=J/kT$ ; where $k$ and $T$ are the Boltzmann constant and the
temperature respectively, and $J$ is the magnetic interaction between spins
$\sigma_{i}$ and $\sigma_{j}$, which can take values $1,2,\dots, q$ for the
$q$-state Potts model.
One of the basic operators, the average cluster size can be defined as
\begin{equation}
ACS = {1 \over N_C} < \sum_{i=1}^{N_C} C_i>
\end{equation}
where $C_i$ is the number of spins in the $i^{th}$ cluster devided by the
total number of spins.
Similar to the energy cumulants, the fluctuations in this quantity
at the transition temperature take a  polynomial form \cite{Borgs:1990} 
\begin{equation}
(<C^2 > - <C>^2)_{max} = A_0 + A_1/L^{d}+ A_2/L^{2d}+\dots.
\end{equation}
where $d$ is the space dimension.

$\;$ 

In the simulation of $7-$state Potts model the measurements are done
on twelve lattices in the range $12 \le L \le 64$. Since the
calculated correlation length $\xi$ of this model is about 50 lattice
sites \cite{Buffenoir:1993}, this set of lattices are considered to be
an indicative set for both very small lattice measurements and for the
lattices of the order of the correlation length. Measurements are done
around the transition temperature and at each temperature $5 \times
10^5$ iterations are performed following the thermalization runs of $5
\times 10^4$ to $10^5$ iterations. For $q=5$ state Potts model, the
measurements are done on lattice $32 \le L \le 128$.  On the largest
three lattices $10^6$ iterations were performed in order to obtain the
same statistical errors.  For all simulation works, cluster update
algorithm ~\cite{Swendsen:1987, Wolff:1989} is employed.

$\;$ 

The first indication towards the realization of our expectations of
the scaling behavior of cluster related operators can be seen in
Fig. 1 where cluster size fluctuation $(FCS)$ data for $q=7$ model is
displayed.  Using only the lattices with $48 \le L$, the cluster size
fluctuations $(FCS)$ data exhibit {\it data collapsing} which is a
manifestation of scaling.  Furthermore, as one can see from Fig. 1,
the data obtained on lattices $28 \le L \le 48$, deviate very little
around the data collapse curve which means a possible correction term
in scaling form has very little effect.  This is an indication that
the asymptotic regime is reached for these lattices.  Such a data
collapse for $L \le 64$ is impossible to observe for the energy
fluctuations, as it has been studied extensively in literature
\cite{Billoire:1992a}.  This observation is also supported by the fits
done to the maxima of the cluster fluctuation data $(FCS)_{max}$. This
quantity is expected to be independent of size in the scaling region
for a strong first-order phase transition. In Fig. 2, the
$(FCS)_{max}$ values are plotted against $1/L^d$. The $(FCS)_{max}$ of
the largest three lattices (L=48, 56 and 64), in most conservative
view, are size independent.  Increasing the number of data points by
considering the next two lattices gives the same constant within
errors and improves the quality of the fit.  The fit can be repeated
by including smaller size lattices $ (L \le 40)$ in which case the
first correction term in Eq. (3) must be included (FORM1 = $A_0 + A_1
/ L^d$).  When all the lattice sizes are included, the results are
$A_0 = 0.0843 \pm 0.0003$ and $A_1 = 0.630 \pm 0.064 $ with $\chi^2 =
0.43$.  The third term in Eq. (3) seems to have no effect on the first
two coefficients and the large error on the coefficient $A_2$
indicates that this term has no room for this set of data.

$\;$ 

In the case of $5-$state Potts model, since the correlation length is
of the order of few thousands of spins, observing asymptotic behavior
by using lattices of size in the range of $L=32$ to $128$ sites can
not be expected. Nevertheless, if one can see the expected finite size
behavior with the correction terms, this can be indicative of the
order of the transition. Fig. 3 shows $(FCS)_{max}$ versus ~$1/L^d$
~for ~$q=5$ ~Potts model. A linear fit in $1/L^d$ ~(FORM1) to the
largest four lattices ($64 \le L$) ~gives ~$A_0=0.0429 \pm 0.0007$
~and ~$A_1=48.1 \pm 5.7$ with $\chi^2=0.16$. For a fit to the smaller
lattices, ~$A_2/L^{2d}$ term becomes necessary (FORM2 = $A_0 + A_1 /
L^d + A_2 / L^{2d}$).  This form fits to the lattices between size 128
down to 56. In this procedure $A_0 \; {\rm and }\; A_1$ remains
unchanged within the error bars.  In order to include smaller lattices
another form (FORM3 = $A_0 + A_1 / L^d + A_2 exp( - A_3 L)$) has also
been tested for all size lattices. This form fits to all of the
lattices (Fig.3.)  while leaving the first coeeficient ($A_0$)
unchanged.

$\;$ 

Recently there has been a new publication in support of the idea
that the global operators carry more information than the local
operators \cite{Oliveira:1995}. The finite size scaling algorithm
based on bulk and surface renormalization of ~de Oliveira
\cite{Oliveira:1992} is tested on q-state Potts models in dimensions
$d=2$ and $3$. The operators chosen are very sensitive to the order of
the transition. They have calculated the magnetic critical index and
obtained with good accuracy the first-order value for even weak
first-order transitions such as for the $d=2$, ~$q=5$ Potts model.
This is in support of our expectations.

$\;$ 

In conclusion, the cluster size fluctuations data for 7-state Potts
model on rather moderate size lattices fits to the expected behavior
of the first-order phase transitions with scaling exponent being the
space dimension.  Measurements on lattices as small as $L=28$ indicate
that a very small correction is needed for scaling of the cluster
fluctuation peaks.  Furthermore, as it is expected from the large
correlation length, the $q=5$ case can not show such good scaling
behavior, but nevertheless the expected form for first-order phase
transition can be observed with very stable fits to the data.  In the
case of energy fluctuations, the data collapse, even for the q=7 state
Potts model case, is seen to be impossible and the lattices considered
failed to give even stable fits.  In this respect it can be emphasized
that the choice of the operators is extremely important, especially in
the case of weak first-order phase transitions.  The results are in
favor of previous observations that the operators which possess global
information on the system can better exhibit the character of the
phase transition.

$\;$

\pagebreak

\section*{Figure captions}

\begin{description}
\item {Figure 1.}  $FCS$ data versus  
                   $(K-K_{c}) \times L^{d}$  for $q=7$      
                   Potts model.

\item {Figure 2.}  $(FCS)_{max}$ plotted as a function of 
                   $1 \over L^{d}$ for $q=7$ Potts model. 
                   The lattice sizes are $12 \le L \le 64$.

\item {Figure 3.}  $(FCS)_{max}$ plotted as a function of 
                   $1 \over L^{d}$ for $q=5$ Potts model. 
                   The lattice sizes are $32 \le L \le 128$.

\end{description}

\pagebreak

\begin{figure}
\psfig{figure=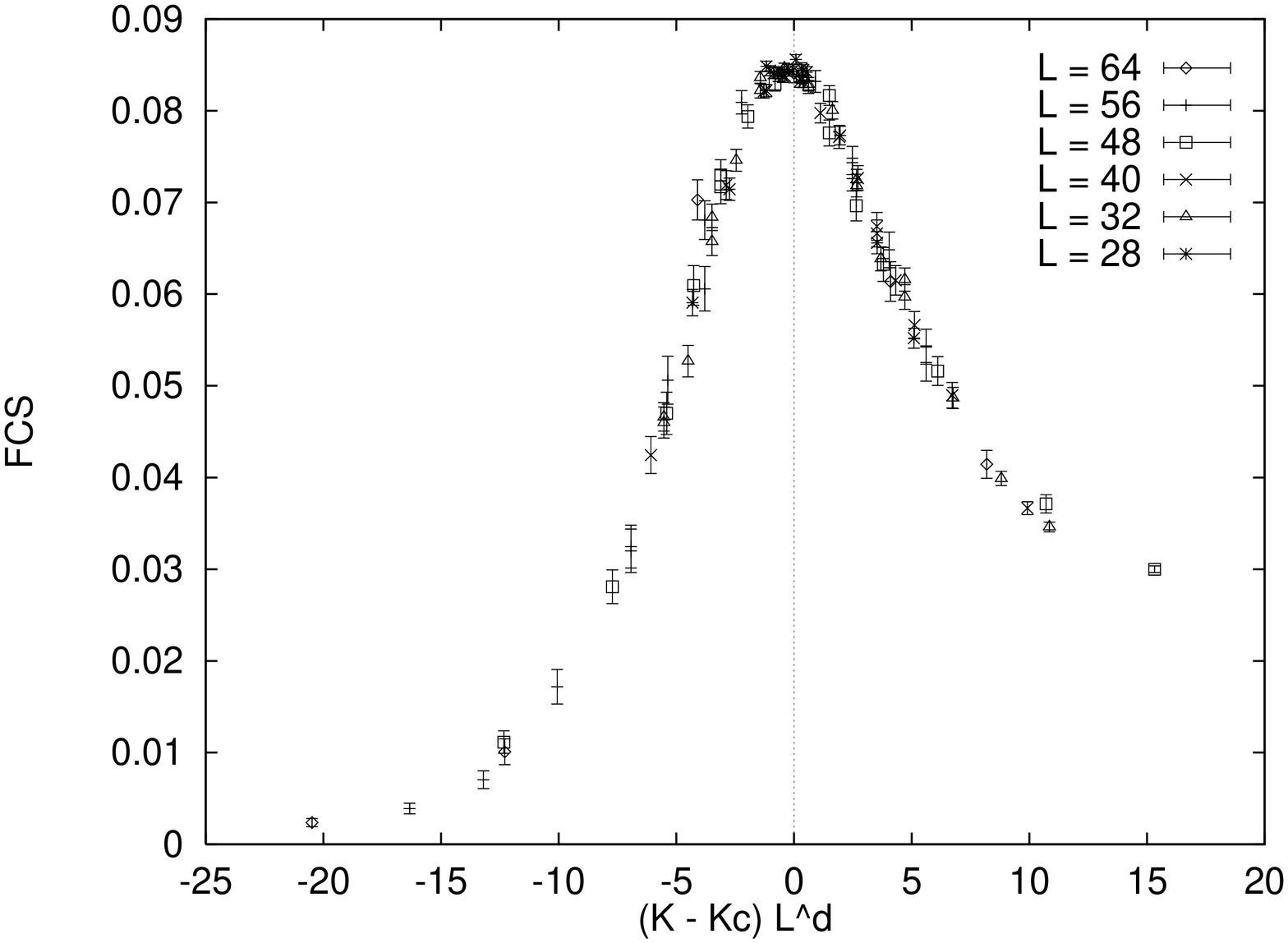,height=8cm,width=12cm,angle=-90}
\caption{}
\end{figure}
\begin{figure}
\psfig{figure=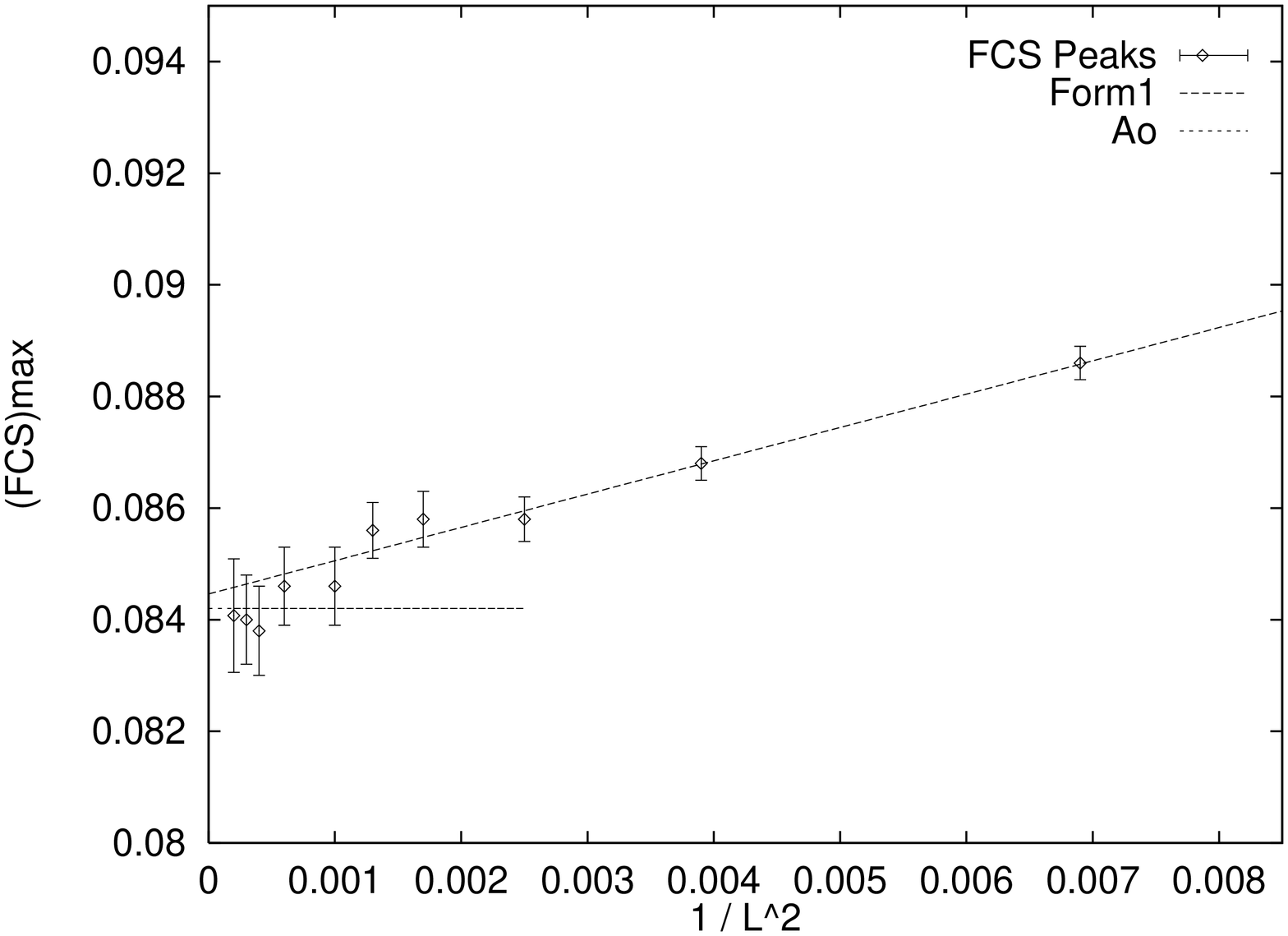,height=8cm,width=12cm,angle=-90}
\caption{}
\end{figure}
\begin{figure}
\psfig{figure=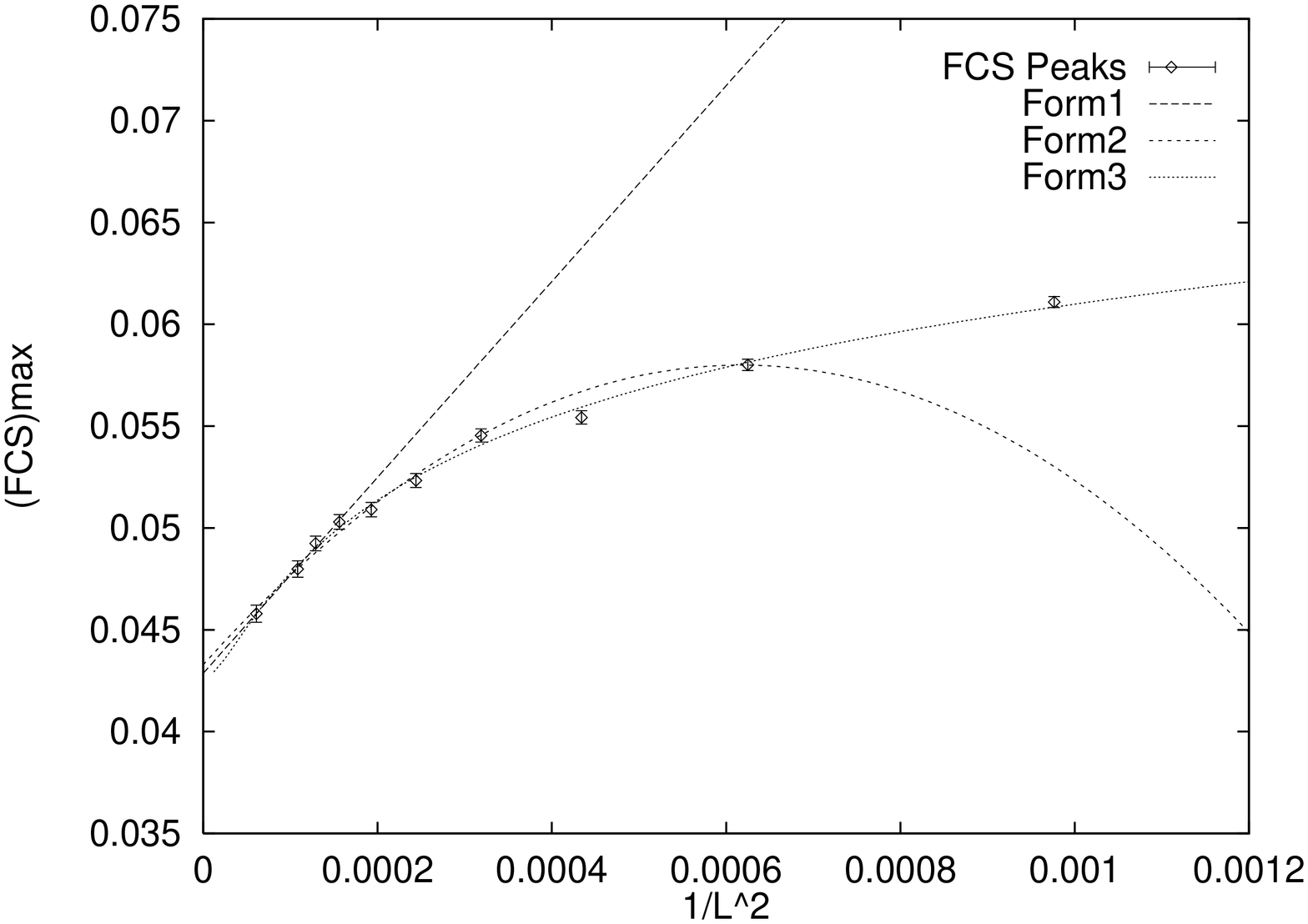,height=8cm,width=12cm,angle=-90}
\caption{}
\end{figure}
\end{document}